\documentstyle[11pt,newpasp,twoside]{article}
\def\edcomment#1{\iffalse\marginpar{\raggedright\sl#1\/}\else\relax\fi}
\reversemarginpar

\begin{document}
\title{Cosmic Matter Distribution: Cosmic Baryon Budget Revisited}
 \author{Masataka Fukugita}
\affil{University of Tokyo, Institute for Cosmic Ray Research,
Kashiwa 277 8582, Japan}

\begin{abstract}
The cosmic baryon budget is revisited using modern observations that
have become available since our first publication. I also present
an estimate for the heavy element abundance.
An increased accuracy in the accounting of the baryon budget reveals  
`missing baryons', which amount to $\approx 35$\% of the 
total. This would provide an interesting test for models of 
the cosmic structure formation.
\end{abstract}

\section{Introduction}

The evolution of the Universe and the formation of cosmic 
structure redistribute
dark matter and baryons. Thus the present day distribution of the matter
reflects the history of the Universe,
and gives us information as to how the cosmic structure formed. For this reason
the matter distribution is taken as a useful constraint  
for models of formation of large-scale structure and galaxies. 
We have published an accounting
of the cosmic baryon budget in 1998 (Fukugita, Hogan \& Peebles 1998,
hereafter FHP).
Since then, much progress has been made in relevant observations,
which include the {\it Sloan Digital Sky Survey} (SDSS: York et al. 2000), 
{\it Wilkinson 
Microwave Anisotropy Probe} (WMAP: Bennett et al. 2003), 
several HI surveys, and others. 
In this talk, I attempt to update the present-day baryon  
budget using modern data, 
and discuss some issues relevant to 
cosmic structure formation. We write the Hubble constant as
$H_0=100h$ km s$^{-1}$Mpc$^{-3}$, but adopt $H_0=72$ km~s$^{-1}$Mpc$^{-1}$
when $h$ is not explicitly denoted.

\section{Baryons in stars}

The basic data used to estimate baryons in stars are the luminosity
function (LF) and the stellar mass to light ratio ($M_s/L$) of galaxies.
The most accurate LF was derived from
the SDSS for five colour bands (Blanton et al. 2001; 2003; Yasuda et al. 
2003).   The first LF from the SDSS given by Blanton et al. (2001) 
is based on earlier
data for the northern equatorial stripe of approximately 200 square
degrees, giving the global luminosity density
${\cal L}_r=(2.58\pm0.28)\times10^8hL_\odot$(Mpc)$^{-3}$.
It turned out, however, that the surface density of galaxies in this region 
for $r<17.9$ mag is somewhat overdense compared to the mean. The
{\it First Data Release} of the SDSS (Abazajian et al. 2003)
now covers 2200 square degrees, and
the LF derived from these data gives a somewhat smaller value
${\cal L}_r=(2.32\pm0.25)\times10^8hL_\odot$(Mpc)$^{-3}$ 
(Yasuda et al. 2003; see also Blanton et al. 2003) in
the $r$ band. The corresponding luminosity density in the $z$ band is
${\cal L}_z=(3.9\pm 0.6)\times10^8hL_\odot$(Mpc)$^{-3}$.

The most extensive analysis for the stellar mass to light ratio is
that by Kauffmann et al. (2003) using $10^5$ SDSS galaxies.
The estimate of the stellar mass using a population synthesis model
depends on metallicity, age, and the star formation history, 
and also on the initial mass function (IMF). 
They estimated the probability distribution of each 
parameter in terms of the Bayesian analysis using five colour photometric
data. In our consideration we limit to bright ($r<15.9$) galaxies,
the median redshift of which is 0.05.
The value of $M/L_z\simeq 1.85$ for luminous galaxies with $M_z<M_z^*-0.8$,
and it decreases gradually to 0.65 for galaxies with $M_z\simeq M_z^*+3$.
The LF-weighted mean is $\langle M/L_z\rangle\simeq1.5$ 
to which a 20\% error is alloted.   

There is still a significant uncertainty in the IMF, 
especially for the subsolar mass. Kauffmann et al. assumed the IMF of
Kroupa et al. (2001). Another typical IMF is that of Reid et al. (1999)
 from DENIS and 2MASS surveys. The former shows an 
increase of the IMF down to $M=0.1M_\odot$, whereas the latter flattens
at $M\leq 1 M_\odot$. This leads to a 40\% difference in the integrated
mass. We take the geometric mean of the two as our central value, allowing
for $\pm20$\% errors. The Kennicutt (1983) IMF gives the integrated mass 
close to our adopted value.
The IMF for $<0.1M_\odot$ is even more uncertain,
but the contribution from this region is small ($<4$\%), provided that the IMF
declines towards smaller masses (e.g., Burgasser et al. 2003). 
The value adopted here 
is 1.10 times that used in FHP, which employed the subsolar mass IMF of
Gould, Bahcall \& Flynn (1996).  The Salpeter IMF, when cut off 
at $0.1M_\odot$, 
gives the integrated mass 1.50 times the adopted value.

Assembling these inputs we obtain 
\begin{equation}
\Omega_{\rm star}=0.0025\pm 0.0008,
\label{eq:star}
\end{equation}
which includes dead stars. This value is 
compared to 0.0019$-$0.0057 of FHP. 
The error arises from the luminosity density 
($\pm15$\%), from $M/L$ ($\pm20$\%) and from IMF ($\pm20$\%), 
which are added in quadrature.

\section{Metal abundance in stars}
\label{sec:metal}

In the analysis of Kauffmann et al. (2003), the metallicity is also
an output, but the results are not available. We here use the
oxygen abundance determined from HII regions for nearby galaxies compiled by 
Kobulnicky \& Zaritsky (1999).
The metallicity shows a correlation with the luminosity of galaxies.
Using oxygen as the metallicity indicator, integration over the 
LF in the $B$ band yields,
 \begin{equation}
Z=10^{6.6\pm 0.15}M_\odot {\rm Mpc}^{-3},
\end{equation}
where the solar composition is assumed. 
The zero point is set by the solar value,
log[O/H]+12 = 8.83 at $(Z/X)_\odot=$0.0230 (Grevesse \& Sauvel 2000), or
$Z_\odot=0.0163$ using $X=0.71$.

There is an additional storage of heavy elements in white dwarfs,
which are liberated only by Type Ia supernovae. From the abundance of
white dwarfs (Bahcall \& Soneira 1980), we estimate the heavy
element (C+O) abundance as
\begin{equation}
Z=10^{7.6}M_\odot {\rm Mpc}^{-3},
\end{equation}
which is much larger than (2). Heavy elements frozen in neutron stars
are also a large amount $\approx 10^{6.9}M_\odot$ Mpc$^{-3}$.

\section{Neutral and molecular gas mass} 

FHP adopted the HI observation of optically selected galaxies by Rao and Briggs
(1993), which yielded $\Omega_{\rm HI}=(2.1\pm 0.6)\times 10^{-4}$ 
at $h=0.72$. Since
then, a number of blind HI surveys were carried out.
Among them the largest sample (1000 galaxies) was obtained by
the HIPASS survey (Zwaan et al. 2003), which gives
$\Omega_{\rm HI}=(4.2\pm 0.7)\times 10^{-4}$, twice higher than the
value of Rao \& Briggs. With the correction for helium
(for both HI and H$_2$), the amount of atomic gas is 
\begin{equation}
\Omega_{\rm HI+HeI}=(6.2\pm 1.0)\times 10^{-4}.
\end{equation}

The molecular hydrogen abundance is estimated from the CO survey of
Keres, Yun \& Young (2003): 
\begin{equation}
\Omega_{\rm H_2}={1.6\pm0.6}\times 10^{-4}.
\end{equation}
This is compared to $\Omega_{\rm H_2}={2.1\pm0.6}\times 10^{-4}$
(FHP) obtained by summing the mean H$_2$ abundance for each morphological
class of galaxies (Young \& Scoville 1991) weighted by the abundance of 
morphologically classified galaxies.

\section{Hot gas in clusters}

In FHP the hot gas abundance in clusters was estimated
by integrating the cluster abundance
for mass $M>1\times 10^{14}hM_\odot$ (Bahcall \& Cen 1993) 
and multiplying the gas fraction
obtained from X ray observations. 
The cluster mass was defined by the Abell radius. Now,
the advancement in cluster studies allows us to
use the mass within $r<r_{200}$, where $r_{200}$ is
the radius at which matter density $\rho=200\rho_{\rm crit}$. 
 From a theoretical ground this may give a better measure for
the mass of the virialised system.
Reiprich \& B\"ohringer (2002) estimated from ROSAT All-Sky Survey
that $\Omega_{\rm cl}=0.012{+0.003\atop-0.004}$ for clusters with
mass larger than $M=4.5\times 10^{13}M_\odot$, which are visible
with X rays. 

The cosmic value of the baryon to total mass ratio from the WMAP 
(Spergel et al. 2003) is 
\begin{equation}
\Omega_b/\Omega_m=0.178(1\pm0.09).
\label{eq:wmapbm}
\end{equation}
We estimate the ratio of the stellar 
to total mass from the mean value of $M/L_B=(450\pm100)h$ and 
$M_s/L_B=4.5(1\pm0.20)$, giving $M_s/M_{\rm tot}=0.014(1\pm 0.30)$.
This is somewhat larger than the stellar mass density 
(1) divided by the total matter density from 
WMAP $\Omega_m=0.26\pm0.05$:
$\Omega_{\rm star}/\Omega_m=0.010\pm0.004$. 
Assuming that the baryon to dark matter ratio in clusters
agrees with the cosmic value and 
subtracting the stellar mass from the total baryonic mass, we estimate the
hot gas abundance:
\begin{equation}
\Omega_{\rm cl~gas}=0.0020\pm0.0006.
\end{equation}
The significant downward shift compared with FHP
is due to the different definition of the radius with which the cluster
mass is defined.

\section{Warm and cool plasma}

FHP inferred the presence of copious warm and cool plasma based on the
universality of the baryon to dark matter ratio at large scales, and 
suggested that this component fills the gap between the cosmic 
baryon abundance from Big Bang nucleosynthesis and
that estimated from {\it observed} baryons in the local Universe. 
The evidence for abundant warm gas around galaxies was presented by 
the detection of O {\tt VI} absorption in the UV spectrum
(Tripp, Savage \& Jenkins 2000).
This year, WMAP gave an accurate estimate for the 
baryon abundance, which agrees with the value from the 
deuterium and helium abundance in Big Bang nucleosynthesis. 
This erases any doubts concerning the
estimate of the cosmic baryon abundance from the nucleosynthesis
argument.

We may estimate the abundance of baryons associated with galaxies 
 from the the mean $M/L$ ratio and the LF of galaxies, assuming that the
baryon to dark matter ratio is universal when averaged over large scales. 
The $M/L$ of Milky Way is known to be $\approx$100 at 200 kpc
(e.g., Kuijken 2003). 
The analysis of
Prada et al. (2003) (see also Zaritsky et al. 1997)
using the motion of 3000 satellites around host galaxies derived from the
SDSS yielded $\langle M/L\rangle =120h$ at the `virial radius'.
The least model-dependent method to measure the mass associated with galaxies
is to use gravitational lensing shear around those with known
redshifts. 
McKay et al. (2001) estimated the galaxy mass using
the SDSS sample, giving $\langle M/L_r\rangle 
=(170\pm21)h$ for $R<260$ kpc from the $r$ band
data. (The $i$ band data  give a smaller value, and $g$ band data give
a larger value.)  

These $M/L$ values are significantly (by about a factor of 2) 
smaller than those for
clusters. Taking the lensing value of $M/L_r=(170\pm20)h$ and the $r$ band 
luminosity density,
we estimate $\Omega_m=0.14\pm 0.02$ for the matter associated with
galaxies (within the virial radius). This leads to 
$\Omega_b=0.025$ when 
multiplied by the universal value of (6). Subtraction 
of $\Omega_s$ and $\Omega_{\rm HI+HeI+H_2}$ gives 
\begin{equation}
\Omega_{\rm w/c~gas}=0.022\pm0.005 
\label{eq:wc1}
\end{equation}
for the warm baryon component around galaxies.

An alternative path to estimate the warm/cool baryon abundance 
is to subtract
stars, neutral and hot ionised gas from the global baryon amount, which 
is accurately known after the WMAP observation:
\begin{equation}
\Omega_{\rm w/c~gas}=0.044-0.0025-0.0020-0.0008=0.039\pm0.004.
\end{equation}

The discrepancy of (8) and (9) implies `missing baryons':
the gap between the two estimates 
suggests the presence of baryons that are not immediately associated with 
galaxies. We do not count in (9) cool ($\approx10^4$K) baryons in  
Lyman $\alpha$ clouds, which were estimated to give $\Omega=0.002\pm0.001$
(FHP), but this contribution is much too small to fill the gap. 
These missing baryons may be in the vicinity of galaxies beyond
a few hundreds of kpc, or associated with dark clumps which do not
shine as galaxies, as they occur in CDM simulations (Ostriker et al. 2003).
The two possibilities may not be necessarily exclusive to each other. 
A possibility is not excluded that the missing baryons are present 
as a highly ionised diffuse component, though this is not very likely
(see below). 

There is also a gap in the dark matter abundance between the cosmic value 
$\Omega_{\rm dm}=0.22$ and the amount associated with galaxies 
$\Omega_{\rm dm}\approx0.12$.
According to the hierarchical clustering calculation,
we anticipate 15\% of baryons are unbound ($M<10^6M_\odot$) and 
25\% are in clumps of mass $10^6M_\odot<M<10^{10}M_\odot$.
CDM simulations (Ostriker et al. 2003) predict a lot of dark, 
low mass clumps.

It would be interesting to note that if we adopt $M/L_r\simeq 320h$ 
($M/L_B\simeq 450h$) of clusters as the universal $M/L$ of galaxies
to estimate the global mass density, we
would obtain $\Omega_m\simeq0.27$ which is consistent with the cosmic
mass density from WMAP. 
This implies
that dark clumps are integrated into clusters, leading to
a large $M/L$ ratio, compared to that for galaxies, for which those
dark components are excluded from accounting. This suggests that
the majority of dark clumps reside in outskirts of galaxies, or in
filaments and groups of galaxies, so that 
dark matter component that is not counted in the estimate of
${\cal L}*\langle M/L\rangle$ is localised, rather than 
smoothly distributed through the Universe, and so are baryons.

\section{Metal abundance outside stars}

Taking the metallicity of interstellar gas given in sect. 2,
we find $Z\approx 1.5\times 10^6M_\odot$. For the
cluster gas,  we infer 
$Z\approx 1.6\times 10^6M_\odot$, adopting 1/3 solar. 
Very little is known for warm and 
hot plasma. If we assume the heavy element abundance of 0.01 solar as in 
globular clusters, or in typical Lyman $\alpha$ clouds, we get  
$Z\approx (0.5-0.9)\times 10^6M_\odot$. Therefore, the metal abundance is
dominated by that in dead stars in galaxies.

\begin{table}[htb]
\begin{center}
\caption{Summary of the cosmic baryon budget}
\vskip5mm
\begin{tabular}{lll}
\tableline
component & FHP & new estimate  \\
\tableline
stars & 0.0019$-$0.0057 & 0.0025$\pm$0.0008\\
HI+HeI gas & 0.00025$-$0.00041 & 0.00062$\pm$0.00010\\
H$_2$ molecular gas & 0.00023$-$0.00037 &0.00016$\pm$0.00006\\
hot plasma in clusters & 0.0014$-$0.0044 &0.0020$\pm$0.0006\\
warm and cold plasma (by sum) & 0.0072$-$0.030 & 0.022$\pm$0.005\\
~~~~~~~~~~~~~~~~~~~~~~~~~~~~ (by subtr.) &      &  0.037$\pm$0.004 \\
\tableline
total & 0.011$-$0.041 & 0.044$\pm$0.004\\
\tableline
\end{tabular}
\end{center}
\end{table}

\section{Summary}

In this report I have presented an updated accounting of the cosmic 
baryon budget, and have given an estimate for the heavy element abundance.
The summary of the cosmic baryon budget is shown in Table 1.
Each entry does not differ greatly from that given in FHP,
but an increased accuracy reveals a missing baryon component 
which amounts to $(0.37-0.22)/0.44\approx35$\% of the total. 
This would provide 
an interesting test for cosmic simulations of the structure formation.

\vskip3mm
I would like to thank Jim Peebles for valuable discussions.  
The work is supported in part by Grant in Aid of the Ministry of Education.

\end{document}